\documentclass[twocolumn,showpacs,preprintnumbers,amsmath,amssymb]{revtex4}
\usepackage{graphicx}
\usepackage{dcolumn}
\usepackage{bm}

\begin{document}
\preprint{}

\title{Period-doubling Instability of Bose-Einstein Condensates\\ Induced in Periodically Translated Optical Lattices}

\author{Nathan Gemelke$^1$, Edina Sarajlic$^1$, Yannick Bidel$^1$, Seokchan
Hong$^2$}
\author{Steven Chu$^3$}
\address{$^1$Department of Physics, Stanford University, Stanford, CA
94305\\$^2$Department of Applied Physics, Stanford University,
Stanford, CA, 94305\\$^3$Directorate, Lawrence Berkeley National
Lab, Berkeley, CA 94720}
\date{\today}

\begin{abstract}
   We observe the formation of momentum
distributions indicative of spatial period-doubling of superfluid
Bose-Einstein condensates in periodically translated optical
lattices. The effect is attributed to dynamic instability of the
condensate wavefunction caused by modulation-induced coupling of
ground and excited bands.
\end{abstract}
\pacs{03.75.Kk,03.75.Lm,32.80.Lg,05.45.Yv}
\maketitle
Atomic Bose-Einstein condensates in optical lattice
potentials have been a focus of many studies in the last decade,
serving as a model system for study of traditionally solid state
phenomena. Interparticle interaction, in combination with the
periodicity of the lattice potential, give rise to interesting
nonlinear dynamics, such as gap soliton formation
\cite{OBERTHALER_gapsoliton} and instabilities
\cite{INGUSCIOdeepinstab,INGUSCIOdyninstab}. These have been
successfully described by the Gross-Pitaevskii equation (GPE),
which describes the mean-field evolution of the condensate
wavefunction. Stationary solutions of the GPE have been shown to
include Bloch waves, as well as non-Bloch states with longer
spatial periodicity \cite{PETHICKperioddouble}, and have been
predicted to be energetically and dynamically unstable for
specific quasimomenta \cite{NIUdyninstab,PETHICKperioddouble}.
Instability occurs when the excitation spectrum of the condensate
is altered by the periodic potential and interaction in such a way
as to make the macroscopic wavefunction unstable to perturbation.
This has been experimentally demonstrated with condensates
accelerated relative to the lattice
\cite{INGUSCIOdyninstab,INGUSCIOdeepinstab}.

We report here on the observation of a period-doubling dynamic
instability introduced by periodically translating (shaking), the
optical lattice potential.  This perturbation couples vibrational
levels in the lattice and creates an effective band structure
through interference of tunneling amplitudes for atoms in
superposed vibrational states. The onset of instability is
detected by directly observing the sudden growth of narrow
momentum components of the condensate wavefunction at half the
lattice recoil momentum, corresponding to perturbation modes which
modulate the condensate wavefunction with a period of two lattice
spacings. The critical onset threshold with shaking amplitude and
frequency is measured and compared to the behavior expected from
calculated band structure and numeric evolution of the GPE.

To observe these effects, samples of $5\times10^4$ bose-condensed
$^{87}Rb$ atoms were prepared in the $|F=2,m_F=2\rangle$ hyperfine
state with no observable thermal component by evaporative cooling
in a triaxial TOP trap \cite{toptrap,toptrap2}. After evaporation,
the magnetic trap frequencies were relaxed to
$(\omega_{qp},\omega_{nqp},\omega_{z}) =2\pi (55,32,28)$Hz, and
the condensate was adiabatically loaded into a set of 1,2 or 3
orthogonal standing waves derived from a 10W, 1.064$\mu m$ Nd:YAG
laser injected by a stable seed laser (Lightwave NPRO model 125).
Two standing waves made a $45^\circ$ angle with the vertical,
while the third was horizontal; each standing wave was formed by
one retroreflected beam focused to 100$\mu m$ at the location of
the condensate, and separated in frequency from the other standing
waves by $80$MHz. All beam intensities were actively stabilized to
$<1\%$ RMS over 2s. Adiabatic loading was achieved by linearly
ramping the standing wave intensities over 400ms. To enable
dynamic translation of the lattice potential, an electrooptic
phase modulator was placed in the retroreflection path of each
standing wave. In this work we used sinusoidal phase modulation
$\phi(t)=\phi_0 \cos(\omega_D t)$, which translated the lattice
potential in the lab frame according to $V_{latt}(x-x_0(t))$,
where $x_0(t)=\lambda \phi_0 /4\pi \cos (\omega_D t)$ is the
modulation amplitude and $\omega_D$ its frequency. We calibrated
the lattice by locating the heating resonance corresponding to
phase modulation at the lattice vibrational frequency
$\omega_{vib}$, where $\hbar \omega_{vib}$ is the energy
difference between the lowest two band centers. When multiple
standing waves were used, their intensities were adjusted to
equalize the vibration frequencies along all directions.
\begin{figure}[t]
\includegraphics[width=3.0 in]{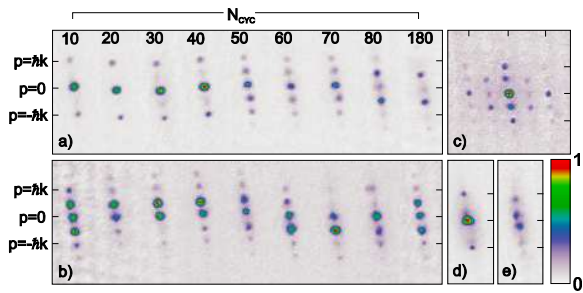}
\caption{\label{fig_1} Absorption images of the atomic
interference following lattice modulation and time-of flight
expansion. (a) Atoms are released at the turning-point of the
lattice translation, after a variable number of modulation cycles,
$N_{cyc}$. For short modulation times, peaks occur only at $p=0$
and $p=\pm \hbar k$. For longer times, peaks develop at $p=\pm
\hbar k/2$. (b) Release occurs within the 76th cycle, spaced by
$\pi /4$ in phase; the center of mass oscillates nearly in phase
with the applied drive. (c) Simultaneous doubling in a 2D lattice
achieved by applying the same modulation as in (a,b) to two
orthogonal standing waves. (d) The effect of modulation above
resonance in a 1D lattice; for small amplitudes, the p=0 peak
develops a shoulder. At larger amplitudes, (e) broad peaks are
apparent between $p=0$ and $p=\pm \hbar k$.  For visibility, (c)
shows atomic absorbtion; (a,b,d,e) show atomic column densities.
Modulation and lattice parameters were (a,b,c)$\omega_{vib}=2\pi$
10.4kHz, $x_0=.04a$, $\omega_D=2\pi$ 7.6kHz, (d)
$\omega_{vib}=2\pi$ 11.0kHz, $x_0=.035a$, $\omega_D=2\pi$ 12.5kHz.
(e) $\omega_{vib}=2\pi$ 11.0kHz, $x_0=.037a$, $\omega_D=2\pi$
12.5kHz.}
\end{figure}

After loading and interrogating condensates in the optical
lattice, the magnetic trap and lattice were switched off in
$200\mu s$, and the atoms were absorbtively imaged \cite{imaging}
after a 30ms time-of-flight. The condensate momentum distribution
was inferred from the matter wave interference patterns of the
released atoms.  Without modulation, we observed interference
peaks at momenta of $n\hbar \mathbf{k}$, where n is an integer,
$\mathbf{k}$ is the inverse lattice vector, indicating
phase-coherent localization in the lattice potential wells. Here,
$\mid \mathbf{k}\mid = 2\pi / a  = 2k_{light}$, where a is the
lattice spacing and $k_{light}$ is the wavevector of the light
forming the optical lattice. The relative peak heights were
determined by the on-site wavefunction of the condensate, whose
center of mass oscillated(Fig.\ref{fig_1}b) when the lattice was
shaken near resonance.

\begin{figure}[tt]
\includegraphics[width=2.5 in]{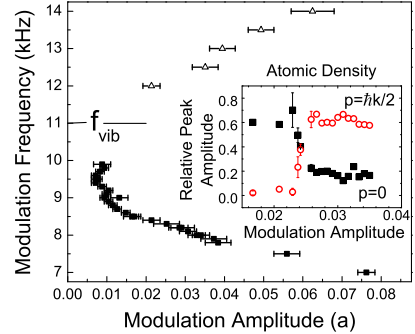}
\caption{\label{fig:ampvdet} Critical drive strengths and
detunings. For $\delta<0$, the period-doubling onset amplitude
decreases near resonance. For $\delta>0$, the p=0 peak develops a
shoulder, and the $p=\pm \hbar k/2$ classes are populated
diffusely; these onsets are marked by open triangles. The inset
shows relative peak atomic density at $|p\,|=0, \hbar k/2$
extracted from absorbtion images as a function of drive amplitude.
Modulation pulse lengths for this data were 10ms.}
\end{figure}

Modulating at a frequency below resonance in a one-dimensional
lattice resulted in the growth of narrow peaks at momenta of $\pm
\hbar k/2$(Fig.\ref{fig_1}a), which also oscillated in phase with
the drive (Fig.\ref{fig_1}b). The presence of peaks spaced by
$\hbar k/2$ indicates periodicity of the condensate wavefunction
on a lengthscale of twice the lattice spacing. These peaks became
visible only beyond a critical frequency-dependent drive amplitude
and following a sufficiently long modulation time. For a typical
lattice depth of 9$E_{rec}$, where $E_{rec}=\frac{\hbar^2
k^2}{8m}$, onset occurred after 5ms of modulation with amplitude
greater than 0.01a at a detuning of
$\delta=\omega_D-\omega_{vib}=-2 \pi$ 2 kHz. The amplitude for
critical onset of this momentum class decreased with decreased
detuning from resonance (Fig.\ref{fig:ampvdet}). Following onset,
peaks were discernable under drive for $\sim 800$ cycles (100ms),
after which they were obscured by a loss of contrast in the
interference pattern. We have found similar behavior at negative
detunings for lattice depths from 3.9 to $19.5 E_{rec}$.

We have investigated the dependence of the critical drive
parameters on the external trapping geometry by varying the
envelope trapping frequencies by $30\%$. This caused no shift in
the onset amplitude at a fixed detuning to 10$\%$. We have also
observed period-doubling in two-dimensional lattices of the same
depth with one or both (Fig.\ref{fig_1}c) axes shaken. Though a
decreased contrast in the interference patterns was observed,
there was no measurable shift in the period-doubling onset
amplitude at a fixed detuning. In three-dimensional lattices, loss
of contrast precluded observation of coherent populations at
$|\mathbf{p}|= \hbar k/2$.

This sudden transition to non-zero momentum states is made
possible by the effect of coupling the ground and first excited
bands via near-resonant translation of the lattice potential.
However, this perturbation acts equally on each lattice site and
cannot itself induce modulations of the wavefunction at twice the
lattice spacing. Interparticle interaction, however, does allow
for spontaneous breaking of this discrete translational symmetry,
and permits coupling between zero and non-zero quasimomentum
states. Accordingly, we consider as candidate mechanisms for this
effect those involving the combined influence of interaction and
lattice modulation.

In the absence of modulation, Bloch states close to the first band
center are stable against decay into other modes
\cite{NIUdyninstab}. Since the second band displays a
comparatively large, inverted tunneling rate, small
modulation-induced admixtures of this state will dramatically
alter the form of the single-particle dispersion, and the
stability of Bloch waves must be reconsidered. For the case of far
off-resonant modulation, band-mixing effects have previously been
described\cite{NIUdynamicinsulator} and
observed\cite{RAIZENdynins} for thermal atoms. This effect becomes
more apparent if we consider the two lowest bands in a 'dressed'
basis set similar to that in \cite{NIUdynamicinsulator}. Here,
each basis state is a time-dependent superposition of delocalized
ground and excited states with the same quasimomentum, chosen to
make the single-particle Hamiltonian time-independent (in a
rotating wave approximation) and diagonal.

\begin{figure}[tt]
\includegraphics[width=3.0 in]{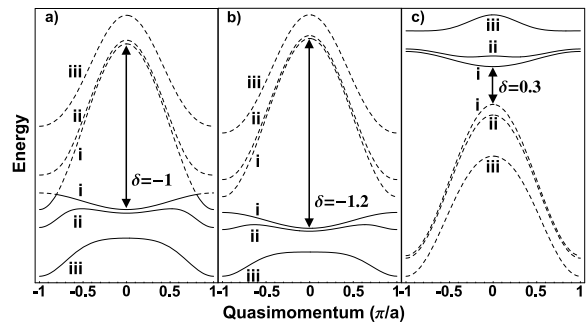}
\caption{\label{fig:bands} Renormalized band structure for
negative (a,b) and positive (c) drive detuning $\delta$, as
calculated by diagonalizing the single particle hamiltonian in the
dressed basis, including only the two lowest bands. Solid (dashed)
lines indicate the energy of the renormalized ground (excited)
band. These were calculated for a lattice depth of 10.5$E_{rec}$,
and inter-band coupling strengths $\hbar \Omega$ of i)0, ii)0.3
and iii)0.9 $E_{rec}$.  Detuning is labeled in units of
$E_{rec}/\hbar$.}
\end{figure}

\begin{figure}[t]
\includegraphics[width=3.0 in]{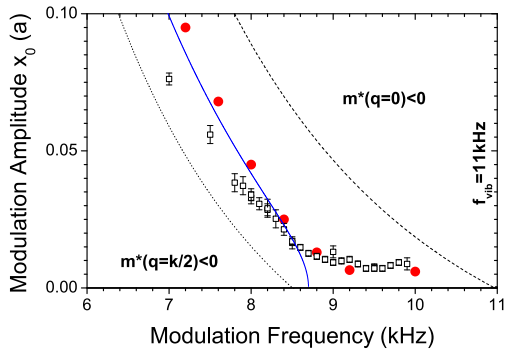}
\caption{\label{fig:numgraph}Comparison of experimental and
numeric data for drive parameters at onset. Boxes show
experimental data points for onset under same conditions as
Fig.\ref{fig:ampvdet}. Solid line shows analytic result for equal
single-particle energy at band center and edge, which occurs for
zero drive amplitude when bands cross. Dashed (dotted) line shows
line of infinite effective mass at band center (edge). Between
dashed and dotted lines effective mass is positive both at band
center and edge. Circles show onset in numeric evolution of the
GPE for $J_0=0.024,J_1=-0.28, U_{00}=U_{10}=(4/3)U_{11}=2\times
10^{-4}$ in units of $E_{rec}$; $\omega_{ext}=2\pi \times 45$Hz
and $N=5\times 10^4$.  Note that the numeric data is a
zero-parameter comparison to experimental data generated from the
measured $\omega_{vib}$.}
\end{figure}

The resulting band structure (Fig. \ref{fig:bands}) can be used to
examine stability of stationary states in the presence of
interaction. As modulation amplitude is increased, the lower
dressed band is flattened by its growing admixture of the bare
excited band, and eventually inverts its curvature at the band
center. In analogy with results for lowest band Bloch wave
stability in the tight-binding limit
\cite{NIUdyninstab,PETHICKperioddouble,MENOTTI_stability}, one
might postulate the emergence of a dynamic instability of the q=0
state near the drive strength sufficient to invert the curvature
of the lower dressed band. This criterion is equivalent to
developing negative effective mass, and may be considered
analogous to the case of attractive interaction with positive
effective mass. This prediction shows a qualitative agreement with
the experimental data (Fig. \ref{fig:numgraph}) for $\delta<0$.

To test this hypothesis, we performed linear stability analysis of
the lattice-periodicity Bloch states for the lower dressed band in
the tight-binding limit of the GPE, following methods in
\cite{NIUdyninstab,PETHICKperioddouble}. Adding a small
perturbation of well-defined momentum to Bloch waves in this band,
we looked for a parameter region where the spectrum of excitations
includes imaginary values, signaling exponential growth or decay
of perturbations to the wavefunction. For negative detunings, one
finds that zero momentum Bloch states first become dynamically
unstable when the modulation amplitude becomes sufficient to make
the energy at the band edge equal to that at the band center
\cite{SARAJLIC_STABILITY_ANALYSIS}. Neglecting higher-order
tunneling contributions, this condition is equivalent to demanding
that the probability for an atom to hop to nearest neighbor sites
is zero, while the hopping rate to next-nearest neighbor sites may
remain nonzero. Close to onset, exponential growth occurs only for
perturbations with momentum close to $\pm \hbar k/2$, consistent
with experimental observations of sharp peaks at that momentum.
The predicted threshold amplitudes for instability are smaller
than those necessary to invert the effective mass, and show a
better quantitative agreement with experimental data (Fig.
\ref{fig:numgraph}) at large negative detunings. This
interpretation fails for small detunings, where the interaction
energy exceeds the dressed band gap and introduces a
non-negligible level mixing.

We note that the curves shown in Fig.\ref{fig:numgraph} were
calculated with knowledge only of the measured lattice vibration
frequency, from which we calculate the unperturbed band structure,
tunneling rates $J_\sigma$ for each band $\sigma$, and Wannier
wavefunctions $\phi_\sigma (x)$. Coupling between bands is modeled
by converting to a frame moving with the lattice, resulting in a
single particle hamiltonian
$H=\frac{p^2}{2m}+V_{latt}(x)+mx\ddot{x}_0(t)$
\cite{NIUdynamicinsulator}. Here the effect of phase modulation is
replaced by an inertial force, which gives an inter-band coupling
strength $\hbar \Omega=m \omega_D^2 x_0 \int d\mathbf{r}\Phi_0^* x
\Phi_1$. The on-site interaction energy is estimated by
approximating the on-site wavefunction transverse to the lattice
$\psi(\mathbf{r_\bot})$ by a Thomas-Fermi profile corresponding to
measured transverse trapping frequencies and and an estimated
average occupation number. The interaction energy for atoms in
bands $\sigma$ and $\sigma '$ is then given by $U_{\sigma \sigma
'}=\frac{4 \pi \hbar^2 a_s(2-\delta_{\sigma \sigma '})}{m}\int
d\mathbf{r} |\Phi_\sigma|^2|\Phi_{\sigma '}|^2 $ , where
$\Phi_\sigma (\mathbf{r})=\psi(\mathbf{r_\bot})\phi_\sigma(x)$,
$a_s$ is the s-wave scattering length, and m the atomic mass. For
simplicity, we approximate the on-site interaction energy
$U_{\sigma \sigma'}$ as independent of site and occupation number;
more accurate treatments may be found in
\cite{SMERZInonlintba,SALASNICHnpse}.

To model the dynamics of the experiment, we have numerically
evolved the GPE in the tight-binding limit for two coupled bands.
Using the approach from \cite{SMERZIsolitons,SALERNOmultiband}, we
describe the condensate wave function with a complex
site-dependent amplitude $c_{j,\sigma}$ for each (bare)
vibrational level $\sigma$, $\langle \mathbf{r} |\Psi \rangle =
\sum_{j} c_{j,\sigma} \Phi_{\sigma} (\mathbf{r}-j \mathbf{a})$. We
substitute $\langle \mathbf{r}|\Psi \rangle$ into the GPE and
numerically evolve the amplitudes $c_{j,\sigma}$ according to
\begin{multline}\label{num_evolution}
    i\hbar \dot{c}_{j,\sigma} = -J_{\sigma} (c_{j+1,\sigma} +
c_{j-1,\sigma}) + V_{j}^\sigma  c_{j,\sigma}
    + \frac{U_{\sigma \sigma}}{2}|c_{j,\sigma}|^2 c_{j,\sigma} \\
    + \frac{U_{\sigma \sigma '}}{4} c_{j,\sigma'}^2 c_{j,\sigma}^*  + \frac{U_{\sigma \sigma '}}{2}|c_{j,\sigma '}|^2
c_{j,\sigma} + \hbar \Omega \cos{(\omega_D t)} c_{j,\sigma '},
\end{multline} where
\begin{equation}
V_{j}^\sigma =  \frac{1}{2}m \omega_{ext}^2 (ja-x_0(t))^2+\hbar
\omega_{vib} \delta_{1,\sigma}+ m\ddot{x}_0(t)ja
\end{equation}
represents the external trapping potential, band energy and the
intra-band effect of the inertial force. We initialize the
calculation with a lower band wavefunction solved to be stationary
in absence of modulation. The chemical potential $\mu$ is chosen
to reproduce the number of atoms N in the experiment. The
calculation was performed with an external potential comparable to
that used in the experiment, and number of sites chosen to safely
exceed the number occupied in the experiment.

The parameters for onset of the $\pm \hbar k/2$ peaks at negative
drive detunings show good agreement with those observed in the
experiment (Fig. \ref{fig:numgraph}), and agree well with the
prediction for instability at the center of the lowest dressed
band described above for large detuning. We have performed the
same numerical evolution for an excited state with a positive
tunneling amplitude; in this case, no period-doubling was
observable.

To investigate the importance of the external trapping potential,
we compare period-doubling onsets with and without $\omega_{ext}$.
To reduce numeric artifacts stemming from the large inertial force
and boundary effects, this comparison was done in the stationary
frame, to first order in the lattice displacement amplitude $x_0$.
Without the external potential, periodic boundary conditions were
used with a site number comparable to the experiment ($40$). In
this case, we find a small initial seed population at $p=\pm \hbar
k/2$ grows rapidly once a sufficient modulation strength is
applied. Without a seed population, observable peaks at $\pm \hbar
k/2$ do not occur for experimentally relevant timescales. With an
external trapping potential this seed population is guaranteed by
the finite extent of the condensate wavefunction, and numerically
results in an observable population at $p=\pm \hbar k/2$ in
several ms, consistent with experimental observations. This
suggests that, while the mechanism for period-doubling exists on a
lattice with discrete translational symmetry, the onset of the
period-doubling population is hastened by an external potential
through its effect on the initial state.

The analysis above applies to modulation performed below
resonance, where the influence of the avoided crossing in
Fig.\ref{fig:bands} is felt most strongly at the band edge.
Experimentally, modulation performed at frequencies higher than
the vibrational resonance first results in the appearance of a
shoulder around the p=0 peak, followed by growth of broad peaks
around $p=\pm \hbar k/2$ (Fig. \ref{fig_1}d,e). The critical drive
amplitude necessary for this behavior is higher than for the
period-doubling onset at an equal detuning of opposite sign (Fig.
\ref{fig:ampvdet}).  While such behavior might be expected from
the band structure in curve ii of Fig.\ref{fig:bands}c, the
details of its onset and dynamics are not yet fully understood.

In conclusion, we have experimentally observed a period-doubling
instability of a $^{87}Rb$ condensate held in a shaken optical
lattice potential, and determined the threshold modulation
conditions necessary for the onset of this behavior. A simple
physical interpretation provides qualitative agreement with the
onset of the observed dynamic instability, while a numerical
simulation of the effect based on the GPE is in quantitative
agreement with this picture.

A new feature in this observation of dynamic instability is the
controlled introduction of instability at zero quasimomentum and
the ability to directly observe the growth of a well-defined
unstable mode. Due to the depth and dimensionality of the lattices
used, it is interesting to consider the effect of fluctuations of
the long range coherence of the condensate near the onset of this
instability. In the dressed-state picture, the modulation-induced
flattening of the lower band at q=0 further increases the already
large ratio of local interaction energy to tunneling amplitude,
pointing to an increasing importance of fluctuations and limited
validity of a mean field description.  In this light, a
description beyond that of the GPE may warrant future study.

We wish to thank Christopher Pethick, Mark Kasevich, Shoucheng
Zhang and Conjun Wu for helpful discussions. This work was
supported in part by grants from AFOSR and the NSF.

\bibliography{finalhkpaper}
\end{document}